\documentclass[aps,prb,superscriptaddress,floatfix,preprint]{revtex4-1}

\usepackage{times}

\usepackage{amsmath}
\usepackage{amsfonts}
\usepackage{amssymb}
\usepackage{epsfig}

\begin{document}

\title{Hong-Ou-Mandel experiment for temporal investigation of single electron fractionalization}

\author
{
V. Freulon$^{1 \dag}$, A. Marguerite$^{1 \dag}$, J.-M Berroir$^{1}$, B. Pla\c{c}ais$^{1}$, A. Cavanna$^{2}$, Y. Jin$^{2}$ and G. F{\`e}ve$^{1}$$^\ast$ \\
\normalsize{$^{1}$ Laboratoire Pierre Aigrain, Ecole Normale Sup\'erieure-PSL Research University, CNRS, Universit\'e Pierre et Marie Curie-Sorbonne Universit\'es, Universit\'e Paris Diderot-Sorbonne Paris Cit\'e, 24 rue Lhomond, 75231 Paris Cedex 05, France}\\
\normalsize{$^{2}$CNRS-Laboratoire de Photonique et de Nanostructures}\\
\normalsize{Route de Nozay, 91460 Marcoussis, France}\\
\normalsize{$^\ast$ To whom correspondence should be addressed;
E-mail:  feve@lpa.ens.fr.} \\
\normalsize{$^\dag$ These authors contributed equally to this work.} \\
}

\begin{abstract}
Coulomb interaction has a striking effect on electronic propagation in one dimensional conductors. The interaction of an elementary excitation with neighboring conductors favors the emergence of collective modes which eventually leads to the destruction of the Landau quasiparticle. In this process, an injected electron tends to fractionalize into separated pulses carrying a fraction of the electron charge. Here we use two-particle interferences in the electronic analog of the Hong-Ou-Mandel experiment in a quantum Hall conductor at filling factor 2 to probe the fate of a single electron emitted in the outer edge channel and interacting with the inner one.  By studying both channels, we analyze the propagation of the single electron and the generation of interaction induced collective excitations in the inner channel. These complementary information reveal the fractionalization process in time domain and establish its relevance for the destruction of the quasiparticle which degrades into the collective modes.
\end{abstract}

\maketitle

Electron/photon analogies have inspired insightful experiments to understand ballistic electronic propagation in quantum conductors using the electronic analog of optical setups \cite{Ji2003, Roulleau2007, Yamamoto2012}. These analogies are based on a single particle picture where electrons, as photons, do not interact with their environment. Although very useful to understand the main features of quantum electronics, this non-interacting picture fails to describe the decoherence mechanisms of single particle excitations \cite{Degiovanni2009} in one dimensional conductors. Recent developments of single-electron emitters \cite{Feve2007, Leicht2011, Fletcher2013, Dubois2013} enable to study these effects at the single particle scale \cite{Adp}, where the transition from single to many-body physics leads to the most drastic changes. Indeed, as Coulomb interaction linearly couples charge densities in the conductor under study and in the nearby ones, it is properly taken into account in terms of the scattering \cite{Safi1999, Degiovanni2010, Hashisaka2013} of charge density waves (plasmons) from the conductor to the environment. As these plasmonic waves are collective excitations involving several electron/hole pairs, Coulomb interaction brings a competition between single electron propagation and the emergence of collective modes. In one dimensional conductors, the emergence of these collective modes leads to the fractionalization \cite{Steinberg2007, Kamata2013} of an injected electron into separated pulses carrying a fraction of the electron charge, causing the destruction of the Landau quasiparticle  \cite{Degiovanni2009, Ferraro2014}.

\begin{figure}[hhhhhh]
 \centerline{\includegraphics[scale=0.85, keepaspectratio]{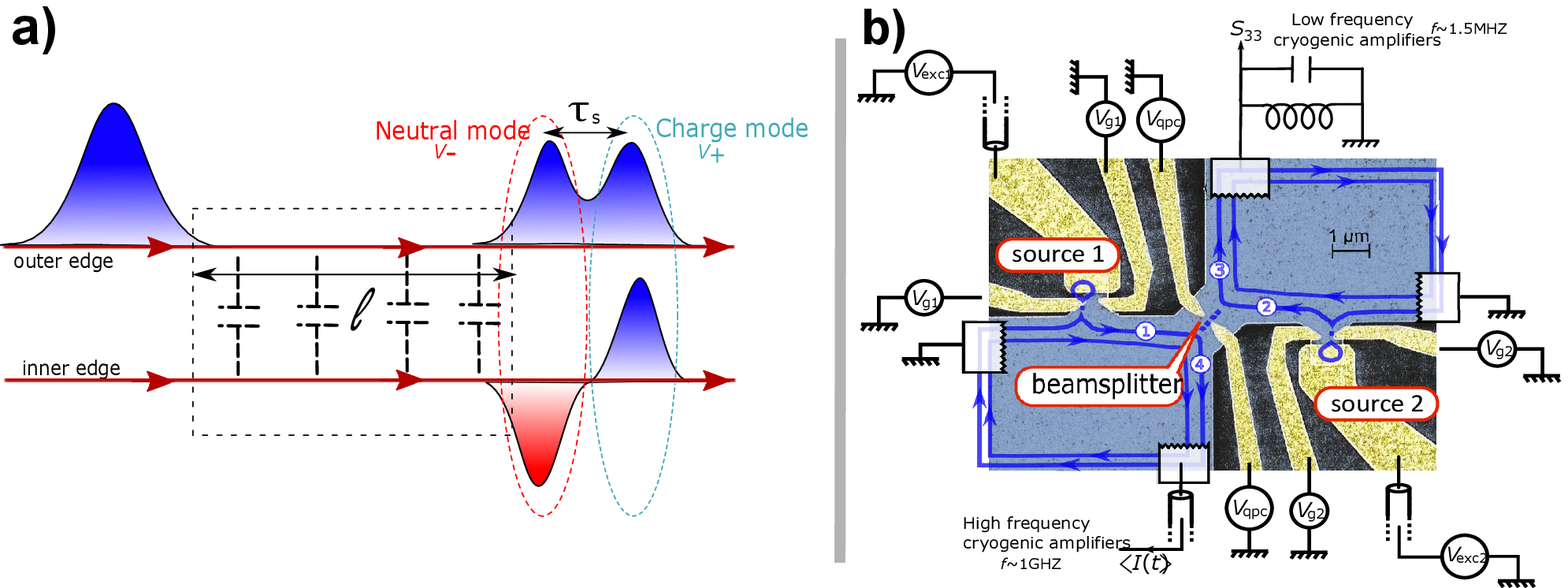}}
 \caption{ \textbf{Sketch of electron fractionalization and measured sample.} \textbf{(a)} Due to interchannel interaction on propagation length $l$, an electronic wave packet emitted on the outer edge channel splits in a charge mode (dashed blue oval) propagating at velocity $v_+$ and a neutral mode (red oval) with velocity $v_-$ separated by time $\tau_s$. The interaction region is represented by a capacitive coupling between the edges in the dashed black box. Negative (positive) charge pulses are represented in blue (red). At the output of the interaction region, the electron on the outer channel has fractionalized in two pulses carrying charge $e/2$. A dipolar current trace has been generated in the inner channel. \textbf{(b)} Modified scanning electron microscope picture of the sample. The electron gas is represented in blue, the edge channels by blue lines and metallic gates are in gold. The emitters are placed at inputs 1 and 2 of the QPC used as an electronic beam-splitter (with a $3$ microns distance between emitter and QPC). Charge emission on the outer channel is triggered by the excitation voltage $V_{\text{exc}, i}$. The dot to edge transmission of source $i$ is tuned by the gate voltage $V_{\text{g}, i}$. The central QPC gate voltage $V_{\text{qpc}}$ can be tuned to partition either the outer or inner edge channel. Average ac current measurements are performed on the splitter output $4$, low frequency noise spectrum measurements $S_{33}$ are performed on output $3$. } \label{Fig1}
 \end{figure}

A two dimensional electron gas in the integer quantum Hall regime at filling factor $\nu=2$ is a suitable testbed to probe the emergence of many-body physics. Firstly, propagation is ballistic and chiral exemplifying electron/photon analogies. Secondly, it enables to probe both the conductor and its environment. At $\nu=2$, charge transport occurs along two co-propagating edge channels carrying opposite spins. The outer channel is the one-dimensional conductor under study, it interacts mainly with the inner channel which provides a well controlled environment, with the possibility to model accurately interchannel interaction. The collective eigenmodes are known \cite{Levkivskyi2008}: for strong interaction, they correspond to the symmetric distribution of charge between channels, called charge mode, propagating with velocity $v_+$ and the antisymmetric distribution of charge called dipolar or neutral mode propagating with velocity $v_-$. As $v_+ \neq v_-$, a single electron wavepacket generated on the outer edge channel propagating on length $l$ splits in two charge pulses \cite{Berg2009} carrying charge $e/2$ (see Figure \ref{Fig1}a) separated by time $\tau_\text{s} =  l/ \bar{v}= l/v_- - l/v_+ \approx  70-80$ picoseconds  (with $\bar{v} \approx 5 \times 10^4$ $\text{m}.\text{s}^{-1}$  and $l\approx 3$ microns from ref. \cite{Bocquillon2013b}). This process is accompanied by the generation of collective excitations in the inner channel with a dipolar current trace : an electron like pulse followed by a hole like one separated by $\tau_\text{s}$ (see Figure \ref{Fig1}a). This mechanism leads to the relaxation and decoherence of the quasiparticle emitted in the outer channel: once the wavepacket has fully fractionalized, the individual electron no longer exists and has been replaced by a cloud of collective excitations \cite{Ferraro2014}.
Interchannel interactions have been revealed by contrast reduction in electronic Mach-Zehnder interferometers \cite{Roulleau2008, Letvin2008} and by relaxation of non-equilibrium energy distribution \cite{LeSueur2010}. The neutral and charge modes have also been observed and characterized measuring high frequency admittance  \cite{Bocquillon2013b} or partition noise \cite{Neder2012, Inoue2014}.

Here, we follow the suggestion of Wahl \cite{Wahl2014} and collaborators and use the electronic analog \cite{Ol'khovskaya2008, Bocquillon2013} of the Hong-Ou-Mandel \cite{HOM} experiment to provide a stringent test of the interaction process. The HOM experiment is based on two-particle interferences occurring through the exchange of indistinguishable particles. First evidenced in quantum conductors with stationary emitters \cite{Liu1997, Samuelsson2004, Neder2007}, two-electron interferences have been recently used to extract information with a few picoseconds resolution \cite{Bocquillon2013, Dubois2013, Grenier2011, Jullien2014} using single electron emitters, in the spirit of the seminal HOM experiment with single photons \cite{HOM}. When indistinguishable fermionic wavepackets collide synchronously on a beamsplitter, they always exit in different outputs, suppressing the random partition noise. Relying on the indistinguishability between input states, two-particle interferometry can be used to compare the temporal traces of the input wavepackets by introducing a delay $\tau$ between emitters emissions. Relying on the interference between two paths, it can also be used to probe the coherence of the inputs states \cite{HOM}.

In this work, we emit a single charge excitation in the outer channel and perform HOM interferometry both on the outer and inner channels after propagation on a 3 microns length. Outer channel interferometry directly probes the fate of the single Landau quasiparticle, inner channel interferometry reveals the collective excitations generated in the interaction process. Combining these complementary informations, we can directly picture the fractionalization in time domain and establish its relevance for the decoherence and destruction of the quasiparticle which degrades into the collective modes.

\section*{Results}

\subsection*{Sample}
The sample is described on Fig. 1b. It is realized in a two dimensional electron gas of nominal density $n_s=1.9\times10^{15}$ $\text{m}^{-2}$ and mobility $\mu=2.4 \times 10^6$ $\text{cm}^2.\text{V}^{-1}.\text{s}^{-1}$ placed a in strong magnetic field $B=4$ T so as to reach a filling factor $\nu=2$ in the bulk. The emitters are two quantum dots synchronously driven by a periodic square excitation applied on the dot top gates with a $40$ ps risetime. They are placed at a distance $l=3.2\pm0.4\,\mu$m (corresponding to the interaction region) of a quantum point contact (QPC) used as the electronic beam-splitter. Source 1 is placed at input 1 of the splitter, source 2 at input 2. Changing the voltage $V_{\text{QPC}}$, the QPC can be set to partition either the outer or the inner edge channel. The dots are only coupled to the outer edge channel such that the current pulse is generated on the outer channel only. The dot to edge transmission $D$ is used to tune the dot emission time and the dot charge quantization. Two configurations are studied: at $D=1$ the dot is perfectly coupled, charge quantization is lost and a classical current pulse (carrying a charge close to $e$) is generated in the outer channel. This configuration provides the shortest emission time and thus the best time resolution. At $D \approx 0.3$, charge is quantized and single quasiparticles are emitted in the outer channel. As we use a periodic square excitation, the electron emission is followed by hole emission \cite{Feve2007} corresponding to the dot reloading, with a repetition time $T=1.10$ ns. The HOM noise $\Delta q (\tau)$  normalized by the random partition noise is measured on output $3$ of the QPC (see Methods).

\subsection*{HOM interferometry reveals fractionalization}
\begin{figure}[hhhhhh]
\includegraphics[width=\columnwidth]{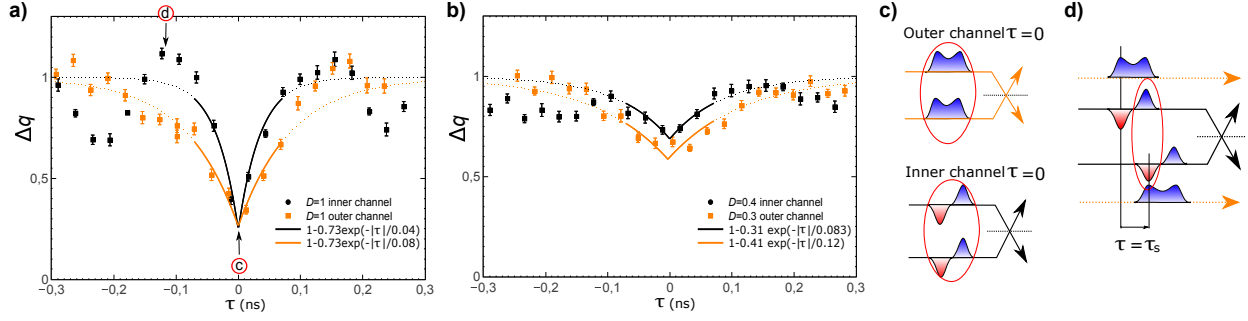}
 \caption{ \textbf{Normalized HOM noise.} \textbf{(a)} $\Delta q(\tau)$ at $D=1$ for outer (orange points) and inner (black points) channel partitioning. Error bars on panels a) and b) equal the standard error of the mean reflecting the statistical dispersion of points. \textbf{(b)} $\Delta q(\tau)$ at $D\approx 0.3$ for outer (orange points) and inner (black points) channel partitioning. Encircled c and d refer to the sketches on panels c) and d). The black and orange dashed lines on both panels represent fits of the dips using the following exponential dependence : $\Delta q(\tau) = 1 - \gamma e^{-|\tau|/\tau_\text{w}}$. The extracted values at $D=1$ are $\gamma = 0.73$ (both for outer and inner channels) and $\tau_\text{w} =40$ ps (inner channel) and $\tau_\text{w} = 80$ ps (outer channel). At $D\approx 0.3$, we have $\gamma = 0.41$ and $\tau_\text{w} =120 $ ps (outer channel) and $\gamma = 0.31$ and $\tau_\text{w} = 83$ ps (inner channel).
 \textbf{(c)} Sketch of current pulses synchronization at $\tau=0$ for the outer and inner channel partitioning. The outer channels are represented as orange lines, the inner as black lines. Negative (positive) charge pulses are represented by blue (red) colors. Pulses colliding synchronously are emphasized by red circles. \textbf{(d)} Sketch of inner and outer channel current pulses when the time delay between the sources is $\tau = \tau_\text{s}$. The inner channels (black lines) are partitioned while the outer ones (orange dashed lines) are not. } \label{Fig2}
\end{figure}

Figure \ref{Fig2} shows $\Delta q(\tau)$ for $D=1$ (Figure \ref{Fig2}a) and $D \approx 0.3$ (Figure \ref{Fig2}b), both when the outer (orange points) or the inner (black points) channels are partitioned. From the outer channel partitioning, we probe the evolution of the generated electron pulse during propagation, inner channel partitioning results from the collective excitations generated by the interaction process.
All the traces show a noise reduction (dip) on short times $\tau$ which is reminiscent of two-particle interference. However significant differences are observed in the width of the HOM dips, labeled $\tau_\text{w}$, which we estimate using an exponential fit. Focusing first on $D=1$: the outer channel dip is roughly twice larger than the inner one : $\tau_\text{w}=80$ ps (outer) versus $\tau_\text{w}=40$ ps (inner). The increased width of the outer channel dip reflects the fractionalization of the current pulse which splits in two pulses of the same sign (see Figure \ref{Fig2}c). The smaller width on the inner channel reflects the dipolar current trace (see Figure \ref{Fig2}c) and equals the temporal extension of the current pulse of a given sign (electron-like or hole like), limited by the excitation pulse rise time. For larger time delays ($|\tau| \approx 100$ ps), the inner channel normalized HOM signal shows an overshoot above unity. As predicted in refs. \cite{Wahl2014, Jonckheere2012}, $\Delta q(\tau) \geq 1$ occur when an electron-like pulse collides with a hole-like one. It occurs in the inner channel for $|\tau| \approx \tau_\text{s}$, the electron part of the inner channel current pulse in input 1 then collides with the hole part of the current pulse in input 2 (see sketch on Figure \ref{Fig2}d). This contrasts with the monotonical increase of $\Delta q(\tau)$ towards 1 for the outer channel. When the dot transmission is decreased to $D=0.3 \pm 0.05$ ($D =0.4 \pm 0.05$ for inner channel partitioning), we observe the expected increase of the  HOM dip width compared to $D=1$, reflecting the increase in the dot emission time: $\tau_\text{w} = 120$ ps (respectively $\tau_\text{w} = 80$ ps) for the outer (respectively inner) channel. Note
that the dot to edge transmission are slightly different for outer ($D \approx 0.3$) and inner ($D \approx 0.4$) channel
partitioning. Due to gate coupling, it is hard to tune the dot transmissions to the exact same values when the
QPC voltage $V_{\text{qpc}}$ is set to partition the outer or the inner channel. This limited accuracy on the dot transmission does not allow for a quantitative comparison between the outer and inner channels dip widths at $D\approx 0.3$.

\subsection*{Decoherence of single electron states}
The contrast $\gamma =1- \Delta q(0)$ measures the degree of indistinguishability between the states at inputs $1$ and $2$, $\gamma =1$ corresponding to full partition noise suppression, $\gamma =0$ to the absence of interference. The contrasts are much higher for $D=1$ ($\gamma \approx 0.73$ for both channels) compared to $D \approx 0.3$ ($\gamma = 0.35$ for the outer channel and $0.25$ for the inner one). This suppression of the contrast is a consequence of interaction induced \cite{Wahl2014} decoherence. In principle, the contrast of the classical pulse ($D=1$) should not be affected by interactions and we attribute the observed reduction (from $1$ to $\approx 0.75$) to residual asymmetries in the colliding pulses. As a matter of fact, when the dot is fully open, a classical charge density wave, or edge magnetosplasmon (EMP), carrying current $I(t)$ is generated in the outer channel, as if it was driven selectively by the time dependent voltage $V(t)= h/e^2 \; I(t)$. The EMP is a collective charge excitation of bosonic nature. It corresponds in the bosonic description, to a product of coherent states : $| \Psi_{\text{in}} \rangle = \bigotimes _{\omega >0 } \big[| \alpha_{\omega} \rangle \otimes | 0 \rangle  \big]$, where the coherent state parameter $\alpha_{\omega}=-I_{\omega}/(e \sqrt{\omega})$  encodes the outer channel current \cite{Degiovanni2009} at pulsation $\omega$ and $|0 \rangle$ is the inner edge in the vacuum state (thermal fluctuations are discarded). As a result from interactions, this EMP is partially transferred to the inner channel at the output of the interaction region \cite{Bocquillon2013b, Grenier2013}: $| \Psi_{\text{out}} \rangle = \bigotimes _{\omega > 0} \big[ |t_{\omega} \alpha_{\omega} \rangle \otimes | r_{\omega} \alpha_{\omega} \rangle  \big]$, $t_{\omega}$ (resp. $r_{\omega}$) is the transmission amplitude to the outer (resp. inner) channel that encodes the interaction parameters. As seen from $| \Psi_{\text{out}} \rangle$, the outer channel (conductor) does not get entangled with the inner one (environment) \cite{Ferraro2014}. A perfect dip $\gamma =1$ should be observed both for the outer and inner channels as long as $I_{1, \omega} =I_{2, \omega}$ and $t_{1, \omega}= t_{2, \omega}$. This can be understood from gauge transformation arguments. Indeed for classical applied voltage pulses $V_1(t)$ and $V_2(t)$, all the applied voltage can be brought to one input only (e.g 2) by the overall shift $-V_1(t)$. Noise is then obviously suppressed ($\gamma =1$) for $V_1(t) = V_2(t)$. The situation is completely different for the single particle state produced at  $D \approx 0.3$. The emission of an electronic excitation with wavefunction $\phi^e(x)$ has no classical counterpart in the bosonic representation and corresponds to a coherent superposition of coherent states  \cite{Degiovanni2009, Ferraro2014}: $| \Psi_{\text{in}} \rangle = \left(\int dx \; \phi^{e}(x) \bigotimes_{ \omega >0} |\lambda_{\omega, x} \rangle  \right) \bigotimes_{\omega  >0}  |0_{\omega} \rangle$ with $\lambda_{\omega, x} =\frac{e^{-i \omega x/v}}{\sqrt{\omega}}$ ($v$ being the Fermi velocity). It gets entangled with the environment after interaction, each coherent state in the superposition leaving a different imprint in the environment: $| \Psi_{\text{out}} \rangle = \int dx \; \phi^{e}(x) \bigotimes _{\omega >0} \big[ |t_{\omega} \lambda_{\omega, x}  \rangle \otimes | r_{\omega} \lambda_{\omega, x} \rangle  \big]$. After tracing out the environment (inner channel) degrees of freedom, outer channel coherence is suppressed, corresponding to a strong reduction of indistinguishability between the inputs, and thus of the interference contrast (the same argument holds for two-particle interferences in the inner channel by tracing on the outer channel degrees of freedom). This suppression shows that, as Coulomb interaction favors the emergence of collective excitations through the fractionalization process, it is accompanied by the progressive destruction of the quasiparticle which degrades into the collective modes \cite{Degiovanni2009, Ferraro2014}.

\subsection*{Comparison between data and model}
\begin{figure}[hhhhhh]
\centerline{\includegraphics[width=\columnwidth, keepaspectratio]{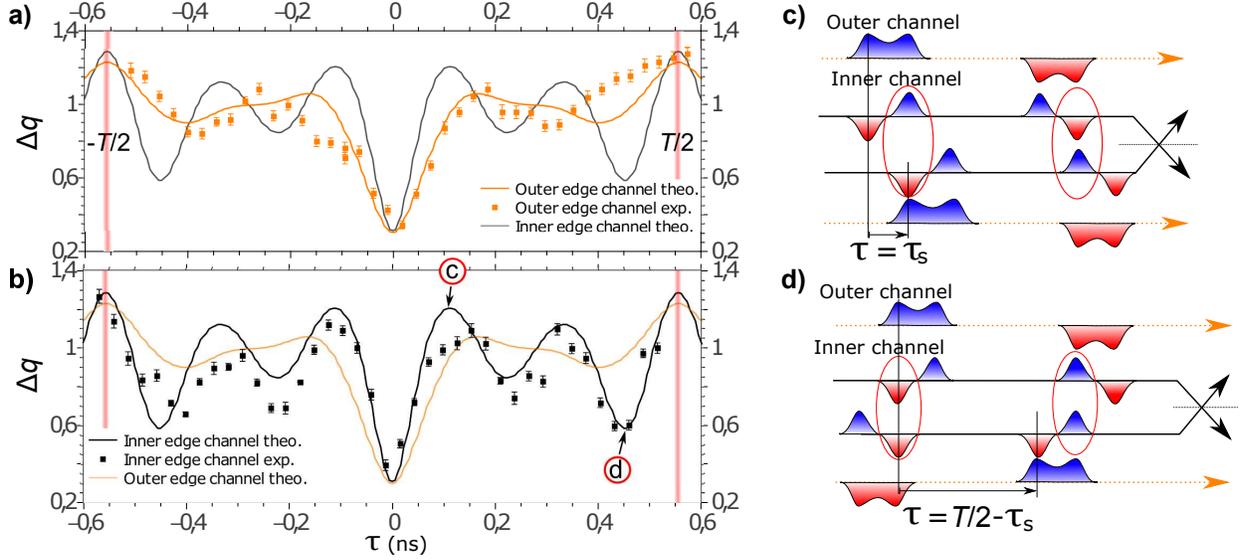}}
 \caption{ \textbf{Temporal investigation of charge fractionalization.} \textbf{(a)} $\Delta q (\tau)$ at perfect dot to edge coupling $D=1$ for outer channel partitioning (orange points). Error bars on both panels equal the standard error of the mean reflecting the statistical dispersion of points. \textbf{(b)} $\Delta q (\tau)$ at $D=1$ for the inner channel partitioning (black points). The orange and black lines on both panels are simulations for $\Delta q(\tau)$. The vertical red lines correspond to a time delay matching the half-period of the excitation drive: $\tau = \pm T/2$. Encircled c and d refer to the sketches on panels c) and d). \textbf{(c)} Sketch of current pulses synchronization at $\tau=\tau_s$ for inner channel partitioning. The outer channels are represented as orange lines, the inner as black lines. Negative (positive) charge pulses are represented by blue (red) colors. Pulses colliding synchronously are emphasized by red circles (electron/hole collision in this case). \textbf{(d)} Sketch of current pulses synchronization at $\tau=T/2 - \tau_s$ for inner channel partitioning. Pulses colliding synchronously are emphasized by red circles (electron/electron and hole/hole collisions in this case).
 } \label{Fig3}
\end{figure}

\begin{figure}[hhhhhh]
\centerline{\includegraphics[width=0.8 \columnwidth, keepaspectratio]{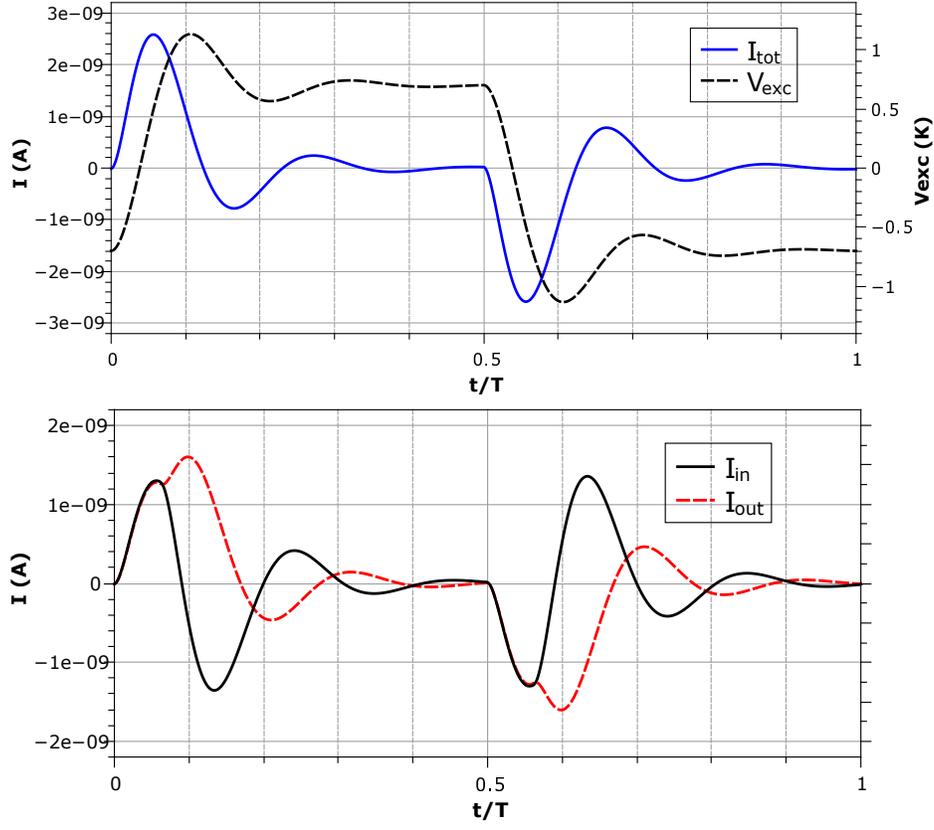}}
 \caption{ \textbf{Output current simulation.} \textbf{(a)} Simulation of the excitation pulse (black dashed line) applied to the dot. The exact shape of the excitation pulse is not known as it is affected by its propagation in the cryostat. The resulting emitted current at $D=1$ before interaction is plotted in blue.  It shows the electron emission sequence followed by the hole one. \textbf{(b)} Outer (red dashed line) and inner (black line) channel currents obtained using $t_{\omega} =  \frac{1+ e^{i \omega \tau_\text{s}}}{2}$ and $r_{\omega}=\frac{1 -e^{i \omega \tau_\text{s}}}{2}$. The outer channel shows the pulse splitting while the inner channel is a dipolar charge excitation.  } \label{Fig4}
 \end{figure}

\begin{figure}[hhhhhh]
\centerline{\includegraphics[width=0.8 \columnwidth, keepaspectratio]{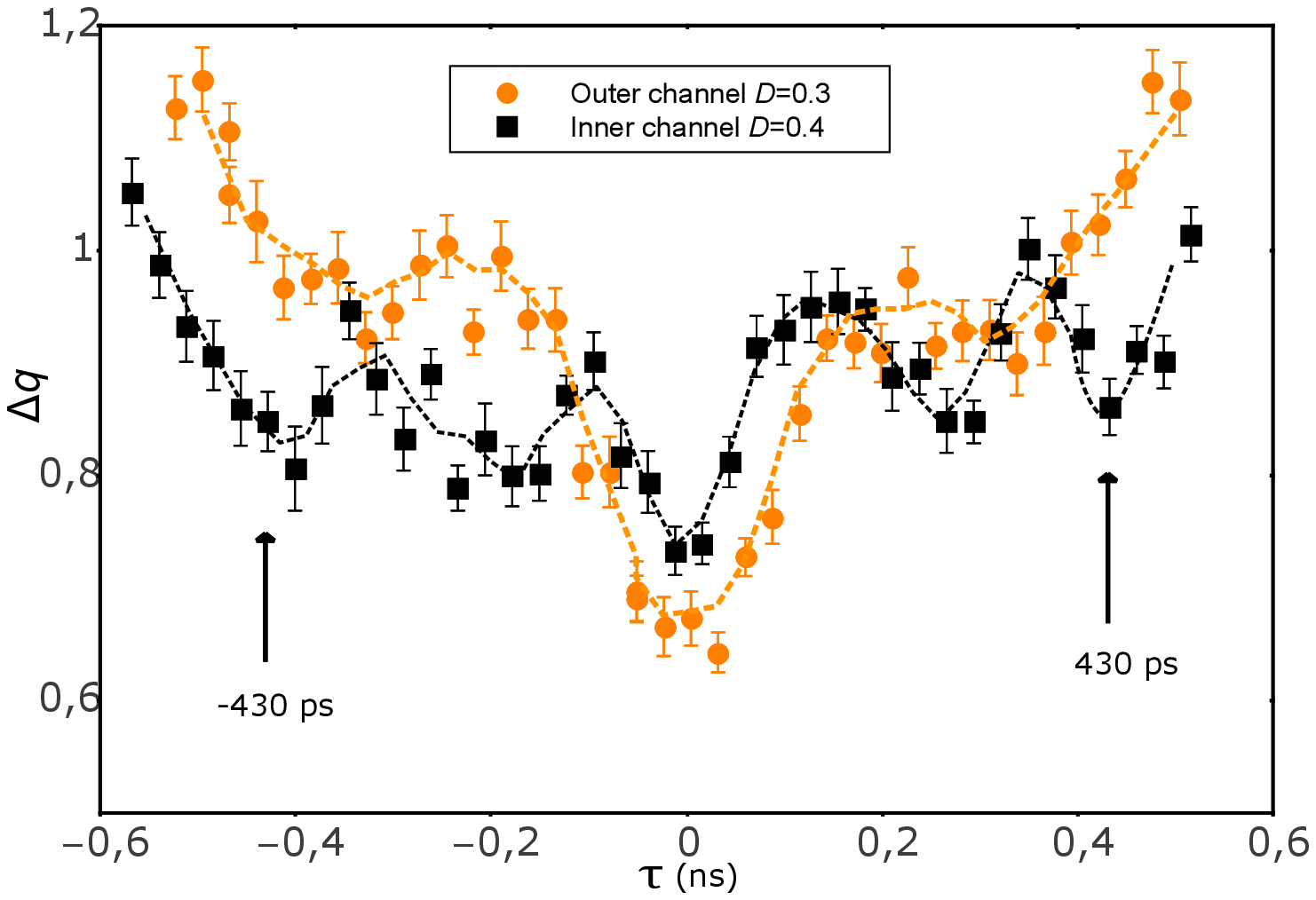}}
 \caption{ \textbf{Full Hong-Ou-Mandel interferometry} $\mathbf{D \approx 0.3}.$ Inner (black points) and outer (orange points) channels normalized HOM noise for the full range of time delays $-T/2 \leq \tau \leq T/2$. Error bars equal the standard error of the mean reflecting the statistical dispersion of points. The orange and black dashed line represent an interpolation of the data points. The arrows represent the position (averaged on the positive and negative values of $\tau$) of the inner channel HOM dip for $\tau \approx T/2 - \tau_\text{s}$.} \label{Fig5}
\end{figure}

Further evidence of fractionalization can be observed on longer time delay $| \tau | \approx T/2$ when electron emission for source 1 is synchronized with hole emission for source 2. For $| \tau | \approx T/2$, $\Delta q(\tau)$ for $D=1$ plotted on Figure \ref{Fig3} exhibits again contrasted behaviors for the outer and inner channels. While it monotonically increases above 1 for the outer channel (see Figure \ref{Fig3}a), as expected for electron/hole collisions, the inner channel shows an additional dip for $|\tau| \approx T/2 -\tau_\text{s}$ (see Figure \ref{Fig3}b). This reveals again the dipolar nature of the inner current: as the dipoles have opposite signs for electron and hole emission sequences, the electron parts of each dipole are synchronized for $|\tau| = T/2 -\tau_\text{s}$ (see sketch on Figure \ref{Fig3}d). A quantitative description of the HOM traces can be obtained (black and orange lines) by simulating the propagation of the current pulse in the interaction region (see Figure \ref{Fig4}) taking interaction parameters $t_{\omega}= \frac{1+ e^{i \omega \tau_\text{s}}}{2}$ and $r_{\omega} = \frac{1-e^{i \omega \tau_\text{s}}}{2}$ and $\tau_\text{s} = l/\bar{v} = 70$ ps  measured on a similar sample \cite{Bocquillon2013b}. The obtained current traces at the output of the interaction region (black and red dashed lines on Figure \ref{Fig4}) reproduce the sketch depicted on Figure \ref{Fig1}a. The good agreement obtained for the HOM trace (Figure \ref{Fig3}b) supports the above qualitative descriptions of the dips observed at $\tau_\text{s}$ and $T/2-\tau_\text{s}$ related to charge fractionalization. Note that an additional spurious modulation of the current resulting from a rebound in our excitation pulse also occur causing an additional dip at $|\tau| \approx 350$ ps on the outer channel and $|\tau| \approx 225$ ps on the inner one.   Finally, Figure $\ref{Fig5}$ presents $\Delta q(\tau)$ at $D\approx 0.3$ for the full range of time shifts $-T/2 \leq \tau \leq T/2$. The qualitative behavior, although strongly blurred by decoherence, is similar to that of Figure \ref{Fig3}. In particular, the additional dip for $|\tau|\approx T/2 - \tau_\text{s}$ is only observed on the inner channel which is a hallmark of single electron fractionalization. Compared to $D=1$, its position is slightly shifted to lower values of $|\tau|$ ($|\tau| \approx 430$ ps), we attribute this difference to the larger width of the emitted current pulse related to the larger emission time.

\section*{Discussion}

We used Hong-Ou-Mandel interferometry to probe single electron coherence on a picosecond timescale and observe single electron fractionalization in two distinct pulses. However, fractionalization goes beyond the mere splitting of a current pulse. Indeed, starting with a single electron state of elementary charge $e$, the final state consists in two pulses of fractional charge $e/2$ and, as such, cannot be described as a single particle state but rather as a collective state composed of several electron/hole pair excitations. The fractionalization process thus results in the destruction of the Landau quasiparticle \cite{Ferraro2014}. Using HOM interferometry, we inferred the quasiparticle desctruction from the decoherence of the electronic wavepacket which results in the suppression of the contrast of two-particle interferences. However, other sources of contrast reduction could be at play in our experiment. The source parameters (transmission $D$ or emission energy) could be different, resulting in the emission of distinguishable wavepackets \cite{Jonckheere2012} and thus of a non-unit contrast. However, given our accuracy, differences in the transmission or in the emission energies cannot explain the contrast reduction we observe. Random fluctuations of the dot energies related to coupling with environmental noise \cite{Iyoda} as well as fluctuations (jitter) in the emission times could also contribute to the contrast reduction.However, even if these contributions cannot be fully discarded, theoretical estimate \cite{Wahl2014} confirm that Coulomb interaction along propagation can explain by itself the contrast reduction we observe.

The $\nu=2$ quantum Hall conductor thus offers a model system to quantitatively study the fractionalization and destruction of the Landau quasiparticle.  Indeed, the environment is well controlled, as the dominant Coulomb interaction results from interchannel interaction. It also offers the possibility to probe simultaneously the the coherence of the emitted state in the outer channel and that of the collective excitation generated in the controlled environment (inner channel). In our analysis, the latter provided the most stringent test of the interaction mechanism as all the signal results from interchannel Coulomb interaction. In particular, the splitting in two distinct pulses can be more easily observed on the inner channel compared to the outer one. To go beyond and characterize fully the single particle decoherence scenario, a quantitative analysis of two-particle interference contrast reduction caused by Coulomb interaction remains to be done.

\section*{Methods}

\subsection*{a.c. current and noise measurements}

The measurements are performed on outputs $3$ and $4$ of the splitter. The ohmic contact on output $4$ is connected to a coaxial line and high frequency cryogenic amplifiers used to measure the average ac current $\langle I(t) \rangle$ generated by the sources and characterize the emitters. The ohmic contact on output $3$ is connected to a resonant circuit (resonant frequency $f_0 \approx 1.5$ MHz) followed by two low-frequency cryogenic amplifiers used to measure the current noise at frequency $f_0$ (see Figure \ref{Fig1}b) after conversion to a voltage noise by the constant impedance $Z=h/ (2 e^2)$ between ohmic contact $3$ and the ground. The average noise power is measured after $1 \times 10^7$ acquisitions in a $78.125$ $\text{kHz}$ bandwidth centered on $f_0$ for a few minutes acquisition time per point. $\Delta q (\tau)$ for each channel are obtained in the following way. We set first the QPC to partition the outer channel (the inner one is fully reflected) and record the random partition noise of each source $\Delta S_{\text{HBT}}^i$ ($i=1, 2$) by measuring the noise difference between the situation where source $i$  is on while source $j$ is off and the situation where both sources are off. Proceeding similarly to measure $\Delta S_{\text{HBT}}^j$, the total random partition noise on the outer channel $\Delta S_{\text{HBT}} = \Delta S_{\text{HBT}}^1 + \Delta S_{\text{HBT}}^2$ is measured. We then proceed to the Hong-Ou-Mandel experiment and measure the noise difference between the situation where both sources are on and the situation where both sources are off. This noise, labeled $\Delta S_{\text{HOM}}(\tau)$, depends on the time difference $\tau$ between the triggering of the two sources, $\tau=0$ corresponding to perfect synchronization. The normalized HOM noise for the outer channel is then defined by $\Delta q (\tau)= \Delta S_{\text{HOM}}(\tau)/\Delta S_{\text{HBT}}$. Setting next the QPC to partition the inner channel (the outer is then fully transmitted), we measure similarly $\Delta q (\tau)$ for the inner channel.

\subsection*{Elements of theory}

The excitation pulse represented on Figure \ref{Fig4} results from a simulation using a step response for the excitation: $V(t) = 0.5 -\frac{e^{-z\omega_\text{n} t}}{\sqrt{1-z^2}} \times  \cos \left( \sqrt{\omega_\text{n}^2(1-z^2)} t+\arcsin(-z)\right)$ for $t \in [0, T/2]$. $z=0.35$ controls the amplitude of the modulation (rebound) and $\omega_\text{n} = 5 \times 2 \pi f$ controls the period of the modulation and the pulse rise time.  The obtained shape is similar to the one observed for our excitation pulse at the top of the cryostat but the parameters $z$ and $\omega_\text{n}$ are different, as the exact shape of the excitation pulse applied at the bottom of the cryostat is not known. The inner and outer channel currents $I_{\text{outer}/\text{inner}}$ (Figure \ref{Fig4}) are computed at the output of the interaction region using EMP scattering parameters $t_{\omega} =\frac{1+ e^{i \omega \tau_\text{s}}}{2}$ and $r_{\omega} = \frac{1-e^{i\omega \tau_\text{s}}}{2}$ corresponding to a short range description of the interaction \cite{Bocquillon2013b}, where $\omega \tau_\text{s} = \omega l/\bar{v} = \omega l \times (1/v_- - 1/v_+)$ is the phase difference between the fast charge and slow neutral modes after propagation length $l$. $\tau_\text{s} = 70$ ps is extracted from the mode dispersion relation measured in ref. \cite{Bocquillon2013b} on a similar sample (coming from the same batch) which established the validity of the short range description for moderate frequencies $f \leq 6$ GHz. The HOM trace are numerically calculated using Floquet scattering formalism \cite{Moskalets2002, Parmentier2012}, driving the outer and inner channels by the excitation $V_{\text{outer}/\text{inner}}(t)= \frac{h}{e^2} I_{\text{outer}/\text{inner}}(t)$. As the same voltages and interaction parameters are used for sources $1$ and $2$, the Floquet simulation predicts a perfect contrast $\Delta q(0)=0$. A finite contrast $\Delta q(0)=0.3$ is thus imposed to the simulated normalized HOM noise. Finally, the unknown parameters $z=0.35$ and $\omega_\text{n} = 5 \times 2 \pi f$ are chosen to reproduce our pulse risetime of $\approx 50$ ps as well as the rebound height in best agreement with the data.

\acknowledgements
We thank F.D Parmentier and F. Pierre for their help in the implementation of the HEMTs for noise measurements, and J. Rech, E. Bocquillon and P. Degiovanni for reading the manuscript. The development of the HEMTs used for cryogenic readout electronics in this experiment was supported in part by the European FP7 space project CESAR grant No. 263455. This work is supported by the ANR grant '1shot', ANR-2010-BLANC-0412.

\section*{Author contributions}
All authors contributed to all aspects of this work.

\newpage

\section*{References}


\begin{thebibliography}{9}




\bibitem{Ji2003}
Y.~Ji, Y.~Chung, D.~Sprinzak, M.~Heiblum, D.~Mahalu, and H.~Shtrikman, \newblock{An electronic Mach-Zehnder interferometer},
\newblock {\em Nature},  \textbf{422}, 415-418 (2003).

\bibitem{Roulleau2007}
P. Roulleau, F. Portier, D. C. Glattli, P. Roche, A. Cavanna, G. Faini, U. Gennser, and D. Mailly, \newblock{Finite bias visibility of the electronic Mach-Zehnder interferometer}, \newblock {\em Phys. Rev. B} \textbf{76}, 161309 (2007).


\bibitem{Yamamoto2012}
M. Yamamoto, S. Takada, C. B\"{a}uerle, K. Watanabe, A.D. Wieck, and S. Tarucha, \newblock{Electrical control of a solid-state flying qubit}, \newblock {\em Nature Nanotechnology} \textbf{7} 247-251 (2012).

 \bibitem{Degiovanni2009}
 P. Degiovanni, C. Grenier and G. F\`{e}ve,\newblock{Decoherence and relaxation of single-electron excitations in quantum Hall edge channels}, \newblock {\em Phys. Rev. B} \textbf{80},
241307(R) (2009).

\bibitem{Feve2007}
G. F\`{e}ve, A. Mah\'{e}, J.-M. Berroir, T. Kontos, B. Pla\c{c}ais, D.C. Glattli, A. Cavanna, B. Etienne and Y. Jin,\newblock{An On-Demand Coherent Single-Electron Source}, \newblock {\em Science} \textbf{316}, 1169-1172  (2007).

\bibitem{Leicht2011}
 C. Leicht, P. Mirovsky, B. Kaestner, F. Hohls, V. Kashcheyevs, E.V. Kurganova, U. Zeitler, T. Weimann, K. Pierz and H.W. Schumacher, \newblock{Generation of energy selective excitations in quantum Hall edge states},  \newblock {\em Semicond. Sci. Technol.} \textbf{26}, 055010
(2011).

\bibitem{Fletcher2013}
J.D. Fletcher, P. See, H. Howe, M. Pepper, S. P. Giblin, J. P. Griffiths, G. A. C. Jones, I. Farrer, D. A. Ritchie, T. J. B. M. Janssen and M. Kataoka, \newblock{Clock-Controlled Emission of Single-Electron Wave Packets in a Solid-State Circuit}, \newblock {\em Phys. Rev. Lett.} \textbf{111}, 216807 (2013).

\bibitem{Dubois2013}
 J. Dubois, T. Jullien, F. Portier, P. Roche, A. Cavanna, Y. Jin, W. Wegscheider, P. Roulleau and D. C. Glattli, \newblock{Minimal-excitation states for electron quantum optics using levitons}, \newblock {\em Nature} \textbf{502}, 659-663 (2013).

\bibitem{Adp}
E. Bocquillon, V. Freulon, F. D. Parmentier, J.-M. Berroir, B. Pla\c{c}ais, C. Wahl, J. Rech, T. Jonckheere, T. Martin, Ch. Grenier, D. Ferraro, P. Degiovanni and G. F\`{e}ve, \newblock{Electron quantum optics in ballistic chiral conductors}, \newblock {\em Annalen der Physik}, \textbf{526},  1-30 (2014).

\bibitem{Safi1999}
 I. Safi, \newblock{A dynamic scattering approach for a gated interacting wire}, \newblock {\em Eur. Phys. J. B} \textbf{12}, 451-455 (1999).

\bibitem{Degiovanni2010}
 P. Degiovanni, Ch. Grenier, G. F\`{e}ve, C. Altimiras, H. le Sueur and F. Pierre, \newblock{Plasmon scattering approach to energy exchange and high-frequency noise in $\nu=2$ quantum Hall edge channels}, \newblock {\em Phys. Rev. B} \textbf{81}, 121302(R) (2010).

\bibitem{Hashisaka2013}
M. Hashisaka, H. Kamata, N. Kumada, K. Washio, R. Murata, K. Muraki, and T. Fujisawa,  \newblock{Distributed-element circuit model of edge magnetoplasmon transport}, \newblock{ \em Phys. Rev. B}, \textbf{88}, 235409 (2013).


\bibitem{Steinberg2007}
H.~Steinberg, G.~Barak, A.~Yacoby, and L.N. Pfeiffer,
\newblock {Charge fractionalization in quantum wires},
\newblock {\em Nature Physics}, \textbf{4}, 116-119, (2007).

\bibitem{Kamata2013}
 H. Kamata,	N. Kumada, M. Hashisaka, K. Muraki and  T. Fujisawa, \newblock{Fractionalized wave packets from an artificial Tomonaga-Luttinger liquid}, \newblock {\em Nature Nanotechnology} \textbf{9}, 177-181 (2014).

\bibitem{Ferraro2014}
D. Ferraro, B. Roussel, C. Cabart, E. Thibierge, G. F\`eve, Ch. Grenier, and P. Degiovanni, \newblock{Real-time Decoherence of Landau and Levitov Quasiparticles in Quantum Hall Edge Channels},  \newblock {\em Phys. Rev. Lett.} \textbf{113}, 166403 (2014).



\bibitem{Levkivskyi2008}
I.P. Levkivskyi and E.V. Sukhorukov, \newblock{Dephasing in the electronic Mach-Zehnder interferometer at filling factor $\nu=2$}, \newblock {\em Phys. Rev. B} \textbf{78}, 045322 (2008).

 \bibitem{Berg2009}
E. Berg, Y. Oreg, E.-A. Kim and F. von Oppen, \newblock {Fractional charges on an integer quantum Hall edge}, \newblock {\em Phys. Rev. Lett.} \textbf{102}, 236402 (2009).


\bibitem{Bocquillon2013b}
E. Bocquillon, V. Freulon, J-.M Berroir, P. Degiovanni, B. Pla\c{c}ais, A. Cavanna, Y. Jin and G. F\`{e}ve, \newblock{Separation of neutral and charge modes in one-dimensional chiral edge channels}, \newblock {\em Nature Communications} \textbf{4}, 1839 (2013).

\bibitem{Letvin2008}
L. V. Litvin, A. Helzel, H.-P. Tranitz, W. Wegscheider, and C. Strunk, \newblock{Edge-channel interference controlled by Landau level filling}, \newblock {\em Phys. Rev. B} \textbf{78}, 075303 (2008).

\bibitem{Roulleau2008}
P. Roulleau, F. Portier, P. Roche, A. Cavanna, G. Faini, U. Gennser, and D. Mailly, \newblock{Direct Measurement of the Coherence Length of Edge States in the Integer Quantum Hall Regime}, \newblock {\em Phys. Rev. Lett.} \textbf{100}, 126802 (2008).

\bibitem{LeSueur2010}
H. Le Sueur, C. Altimiras, U. Gennser, A. Cavanna, D. Mailly, and F. Pierre \newblock{Energy Relaxation in the Integer Quantum Hall Regime}, \newblock {\em Phys. Rev. Lett.} \textbf{105}, 056803 (2010).


\bibitem{Neder2012}
I. Neder, \newblock{Fractionalization noise in edge channels of integer quantum Hall states}, \newblock{\em Phys. Rev. Lett.}, \textbf{108}, 186404 (2012).


\bibitem{Inoue2014}
H. Inoue, A. Grivnin, N. Ofek, I. Neder, M. Heiblum, V. Umansky, and D. Mahalu, \newblock{Charge Fractionalization in the Integer Quantum Hall Effect}, \newblock {\em Phys. Rev. Lett.}  \textbf{112}, 166801 (2014).

\bibitem{Wahl2014}
C. Wahl, J. Rech, T. Jonckheere and T. Martin, \newblock{Interactions and charge fractionalized in an electronic Hong-Ou-Mandel interferometer}, \newblock {\em Phys. Rev. Lett.} \textbf{112}, 046802 (2014).

\bibitem{Ol'khovskaya2008}
S. Ol'khovskaya, J. Splettstoesser, M. Moskalets, and
M. B\"{u}ttiker, \newblock{Shot noise of a mesoscopic two-particle collider}, \newblock {\em Phys. Rev. Lett.} \textbf{101}, 166802 (2008).

 \bibitem{Bocquillon2013}
E. Bocquillon, V. Freulon, J.-M. Berroir, P. Degiovanni, B. Pla\c{c}ais, A. Cavanna, Y. Jin, and G. F\`{e}ve, \newblock{Coherence and indistinguishability of single electrons emitted by independent sources}, \newblock {\em Science} \textbf{339}, 1054-1057 (2013).

\bibitem{HOM}
C. K. Hong, Z. Y. Ou, and L.Mandel, \newblock{Measurement of subpicosecond time intervals between two photons by interference}, \newblock {\em Phys. Rev.
Lett.} \textbf{59}(18), 2044–2046 (1987).

\bibitem{Liu1997}
R. C. Liu, B. Odom, Y. Yamamoto, and S. Tarucha, \newblock{Quantum interference in electron collision}, \newblock {\em Nature} \textbf{391},  263-265 (1997).

\bibitem{Samuelsson2004}
 P. Samuelsson,  E. V. Sukhorukov and M. B\"{u}ttiker, \newblock{Two-Particle Aharonov-Bohm Effect and Entanglement in the Electronic Hanbury Brown–Twiss Setup}, \newblock {\em Phys. Rev. Lett.} \textbf{92}, 026805 (2004).

\bibitem{Neder2007}
I. Neder, N. Ofek, Y. Chung, M. Heiblum, D. Mahalu, and V. Umansky, \newblock{Interference between two indistinguishable electrons from independent sources}, \newblock {\em Nature} \textbf{448}, 333-337 (2007).

\bibitem{Grenier2011}
C. Grenier, R. Herve, E. Bocquillon, F. D. Parmentier, B. Pla\c{c}ais, J. M. Berroir, G. F\`{e}ve, and P. Degiovanni, \newblock{Single-electron quantum tomography in quantum Hall edge channels}, \newblock{\em New Journal of physics} \textbf{13}, 093007 (2011).

\bibitem{Jullien2014}
T. Jullien, P. Roulleau, B. Roche, A. Cavanna, Y. Jin, and D. C. Glattli, \newblock{Quantum tomography of an electron}, \newblock {\em Nature} \textbf{514}, 603-607 (2014).


\bibitem{Jonckheere2012}
T. Jonckheere, J. Rech, C.Wahl, and T. Martin, \newblock{Electron and hole Hong-Ou-Mandel interferometry}, \newblock {\em Phys. Rev. B} \textbf{86}, 125425 (2012).

\bibitem{Grenier2013}
Ch. Grenier, J. Dubois, T. Jullien, P. Roulleau, D.C. Glattli, P. Degiovanni, \newblock{Fractionalization of minimal excitations in integer quantum Hall edge channels}, \newblock {\em Phys. Rev. B} \textbf{88}, 085302 (2013).

\bibitem{Iyoda}
E. Iyoda, T. Kato, K. Koshino, T. Martin, \newblock{Dephasing in single-electron generation due to environmental noise probed by Hong Ou Mandel interferometry},
\newblock{\em Phys. Rev. B} \textbf{ 89}, 205318 (2014).


\bibitem{Moskalets2002}
M. Moskalets and M. B\"{u}ttiker, \newblock{Floquet scattering theory of quantum pumps}, \newblock{\em Phys. Rev. B} \textbf{66}, 205320 (2002).

\bibitem{Parmentier2012}
F.D. Parmentier, E.~Bocquillon, J.-M. Berroir, D.C. Glattli, B.~Pla\c{c}ais,
  G.~F\`{e}ve, M.~Albert, C.~Flindt, and M.~B\"{u}ttiker,
\newblock {Current noise spectrum of a single particle emitter: theory and
  experiment},
\newblock {\em Phys. Rev. B}, \textbf{85}, 165438, (2012).





\end{thebibliography}
\end{document}